\documentclass[11pt,twocolumn,aps,prb,superscriptaddress,reprint,citeautoscript,sort&compress,square]{revtex4-1}

\usepackage{graphicx,amsmath,amssymb,color}
\usepackage{tabularx, ctable}
\usepackage{ulem}
\mathchardef\mhyphen="2D

\usepackage{lineno}
\usepackage{upgreek}

\usepackage{numprint}

\newcommand{\SI}[1]{\textcolor{black}{#1}}

\usepackage[breaklinks=true,colorlinks=false,citecolor=blue,linkcolor=blue,urlcolor=blue] {hyperref}

\setcitestyle{super}

\newcommand{\mum}{\,\mathrm{\upmu m}}

\begin{document}

\title{Single-photon detection using large-scale high-temperature MgB$_2$ sensors at 20\,K }

\author{I. Charaev$^{*}$}
\affiliation{Massachusetts Institute of Technology, Cambridge, MA 02139, USA}
\affiliation{University of Zurich, Zurich 8057, Switzerland}
\author{E. K. Batson}
\affiliation{Massachusetts Institute of Technology, Cambridge, MA 02139, USA}
\author{S. Cherednichenko$^{*}$}
\affiliation{Department of Microtechnology and Nanoscience, Chalmers University of Technology, Göteborg SE-41296, Sweden}
\author{K. Reidy}
\affiliation{Massachusetts Institute of Technology, Cambridge, MA 02139, USA}
\author{V. Drakinskiy}
\affiliation{Department of Microtechnology and Nanoscience, Chalmers University of Technology, Göteborg SE-41296, Sweden}
\author{Y. Yu}
\affiliation{Raith America, Inc., 300 Jordan Road, Troy, NY 12180}
\author{S. Lara-Avila}
\affiliation{Department of Microtechnology and Nanoscience, Chalmers University of Technology, Göteborg SE-41296, Sweden}
\author{J. D. Thomsen}
\affiliation{Massachusetts Institute of Technology, Cambridge, MA 02139, USA}
\author{M. Colangelo}
\affiliation{Massachusetts Institute of Technology, Cambridge, MA 02139, USA}
\author{F. Incalza}
\affiliation{Massachusetts Institute of Technology, Cambridge, MA 02139, USA}
\author{K. Ilin}
\affiliation{Institute of Micro- and Nanoelectronic Systems, Karlsruhe Institute of Technology (KIT), 76187 Karlsruhe, Germany}
\author{A. Schilling}
\affiliation{University of Zurich, Zurich 8057, Switzerland}
\author{K. K. Berggren*}
\affiliation{Massachusetts Institute of Technology, Cambridge, MA 02139, USA}

\begin{abstract}
\textbf{
Ultra-fast single-photon detectors with high current density and operating temperature can benefit space and ground applications, including quantum optical communication systems, lightweight cryogenics for space crafts, and medical use. Here we demonstrate magnesium diboride (MgB$_2$) thin-film superconducting microwires capable of single-photon detection at 1.55\,$\mum$ optical wavelength. We used helium ions to alter the properties of MgB$_2$, resulting in microwire-based detectors exhibiting single-photon sensitivity across a broad temperature range of up to 20\,K, and detection efficiency saturation for 1~$\mum$ wide microwires at 3.7\,K. Linearity of detection rate vs incident power was preserved up to at least ~100 Mcps. Despite the large active area of up to 400$\times$400~$\mu$m$^2$, the reset time was found to be as low as $\sim1$\,ns. Our research provides new possibilities for breaking the operating temperature limit and maximum single-pixel count rate, expanding the detector area, and raises inquiries about the fundamental mechanisms of single-photon detection in high-critical-temperature superconductors.}

\begin{center}
\end{center}

\end{abstract}
\maketitle
 Superconducting Nanowire Single-Photon Detectors (SNSPDs)\cite{semenov2001quantum} have become crucial for a variety of applications, including quantum optics\cite{Hadfield2009,SPD4Qcomp1,SPD4Qcomp2,SPD4Qcomp3} and security\cite{Hadfield2009,SPD4Qcomm1,SPD4Qcomm2,SPD4Qcomm3}, deep-space communication\cite{10.1117/12.2307388}, biomedical imaging\cite{zhao2017single,xia2021short,ozana2021superconducting}, and light detection and ranging (LIDAR)\cite{GUAN2022107102, Hu:21,Salvoni_2019}. SNSPDs have achieved record-breaking efficiencies: a broadband high-efficiency detection from the soft X-ray\cite{doi:10.1063/1.4759046} to mid-infrared spectral range\cite{doi:10.1063/5.0048049}; close to 100 \% system detection efficiency~\cite{reddy2020superconducting,Hu20,Chang}; sub-3\,ps temporal resolution (timing jitter)\cite{korzh2020demonstration}; and extremely low noise (referred to as dark count rate) of 6 × 10$^{-6}$ cps\cite{hochberg2019detecting}. Despite this significant progress, the detection system requires cooling to at least 4.2\,K (and often lower) due to the low critical temperature, $T_\text{C}<$ 10\,K, of utilized superconductors (NbN, NbTiN, WSi, MoSi)\cite{Natarajan_2012}. These temperatures require costly and bulky cryocoolers that limit practical application. As a consequence, there has been a significant interest in superconductors with higher critical temperatures that would allow for operation at elevated temperatures~\cite{shibata2021review}.

In addition to the critical temperature, another material-related limitation arises due to the kinetic inductance, $L_{\text k}$, of superconducting devices\cite{doi:10.1063/1.2183810}. In order to achieve a high photon-detection efficiency, one typically uses superconducting films of thicknesses $\leq$ 10 nm\cite{hofherr2010intrinsic} and high-resistivity (disordered) superconductors\cite{doi:10.1063/5.0048049}. As a result, the films exhibit extremely large values of $L_{\text k}$ ($10^2$-$10^3$ pH/square) thereby limiting the detector reset time\cite{kerman2006kinetic}, particularly problematic for large-area SNSPDs. The high critical temperature ($T_\text{C}$) superconductors with low kinetic inductance may be regarded as viable alternatives for detectors.

\begin{figure*}[ht!]
	\centering\includegraphics[width=1\linewidth]{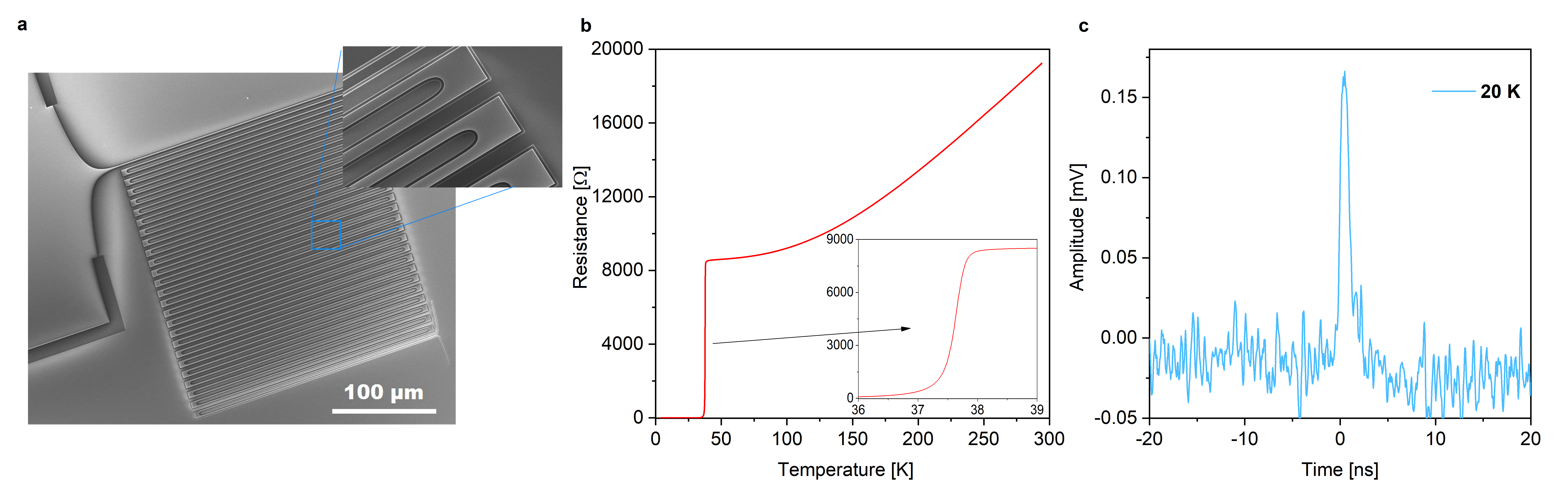}
	\caption{\textbf{Superconducting MgB$_2$ microwire-based single-photon detectors.} \textbf{a,}  Scanning-electron microscopy (SEM) image of a 1 $\mu$m-wide  meander-shaped microwire device. The smooth transitions of the microwires to the electrodes prevent current crowding. The scale bar is 100~$\mu$m. \textbf{b,} Example of the $R(T)$ dependence of a 5 $\mu$m-wide
 meander-shaped microwire device. The measurement was done in a two-terminal configuration. The residual resistance ratio ($RRR$) was 2.25.  \textbf{c,} The voltage pulse from a photon event measured in the 1~$\mum$ MgB$_2$ microwire single-photon detector at 20\,K and $\lambda=1.5\mum$. The device was biased to 98$\%$ of their switching current.} 
	\label{Fig1}
\end{figure*}

In an attempt to increase the operational temperature, higher $T_\text{C}$ materials have been widely explored. Early studies of thin films have suffered from serious challenges\cite{Ejrnaes_2017} in the development of quantum detectors based on high-$T_\text{C}$ superconductors. The lateral proximity effect\cite{charaev2017proximity} with internal constrictions and grain boundaries\cite{dane2017bias}, non-uniform distribution of the superconducting order parameters across the structure\cite{kimmel2019edge}, instability of superconducting parameters\cite{singh2013spatial}, sensitivity to baking, and chemical interaction during the fabrication all impose limitations on the capacity to replicate detectors in a consistent manner. Moreover, the reduction of sensor dimensions, especially crucial for the detection of infrared photons with low energy, may lead to a suppression of superconductivity in low- and high-$T_\text{C}$ materials\cite{arpaia2017transport}. This phenomenon could limit both the sensitivity and operating temperature of potential devices. 


Very recently, attempts to realize SNSPDs using two-dimensional high-$T_\text{C}$ materials revealed single-photon response in superconducting nanowires made of thin flakes of Bi$_2$Sr$_2$CaCu$_2$O$_{8+\delta}$, with single-photon response up to 25\,K \cite{Charaev2023,Merino_2023}. One major drawback of exfoliated flakes, however, is the inability to scale up the detector's active area. 

Magnesium diboride (MgB$_2$) is a promising candidate for single-photon detectors that has been pursued previously for SNSPDs with mixed success\cite{shibata2013fabrication,Cherednichenko_2021,velasco2016high}. The low kinetic inductance ($L_{\text k}$$\sim$2\,pH/square\cite{Cherednichenko_2021}) and critical temperature of 39\,K make MgB$_2$ attractive from practical viewpoints. In SNSPDs made of MgB$_2$ thin films with $T_\text{C}=30~$K, single-photon response in the optical range has been demonstrated at $T\approx 10~$K\cite{shibata2013fabrication,velasco2016high} and at 5K for the near-infrared\cite{10.1063/1.3518723,Cherednichenko_2021}. However, the non-uniformity of thin films over large areas\cite{wolak2015fabrication}, relatively low critical temperature\cite{shibata2013ultrathin}, and low switching current in MgB$_2$ thin films remain essential unsolved challenges. The latest progress in Hybrid Physical-Chemical Vapor Deposition (HPCVD) has illuminated a route to grow high-quality superconducting films of MgB$_2$ for single-photon detectors.

In this work, we demonstrate SNSPDs fabricated from high-quality MgB$_2$ films with an active area of hundreds of square micrometers, critical temperature approaching the bulk value, and single-photon sensitivity up to 20\,K (Fig.\ref{Fig1}). To achieve this result, we used a helium-ion-beam-based irradiation process~\cite{PhysRevApplied.12.044040} that additionally enabled saturated detection of $1.55\mum$ single photons at 3.7\,K. In contrast to previous MgB$_2$ detector attempts, we chose the micro-scale SNSPD geometry (width of the wire, $W$ $\approx$ 1 $\mu$m) recently demonstrated in a number of reports using low-$T_\text{C}$ superconductors\cite{chiles2020superconducting,charaev2020large}. In comparison to nanowires ($W$ $\approx$ 100\,nm), superconducting microwires have far larger switching currents and a lower total kinetic inductance, making them attractive for the fabrication of large-area detectors. We observed a reset time of $\sim 1\,\text{ns}$ enabling a count rate of 100 Mcps for a device with an active area up to 400$\times$400~$\mu$m$^2$. 

\begin{figure*}[ht!]
	\centering\includegraphics[width=1\linewidth]{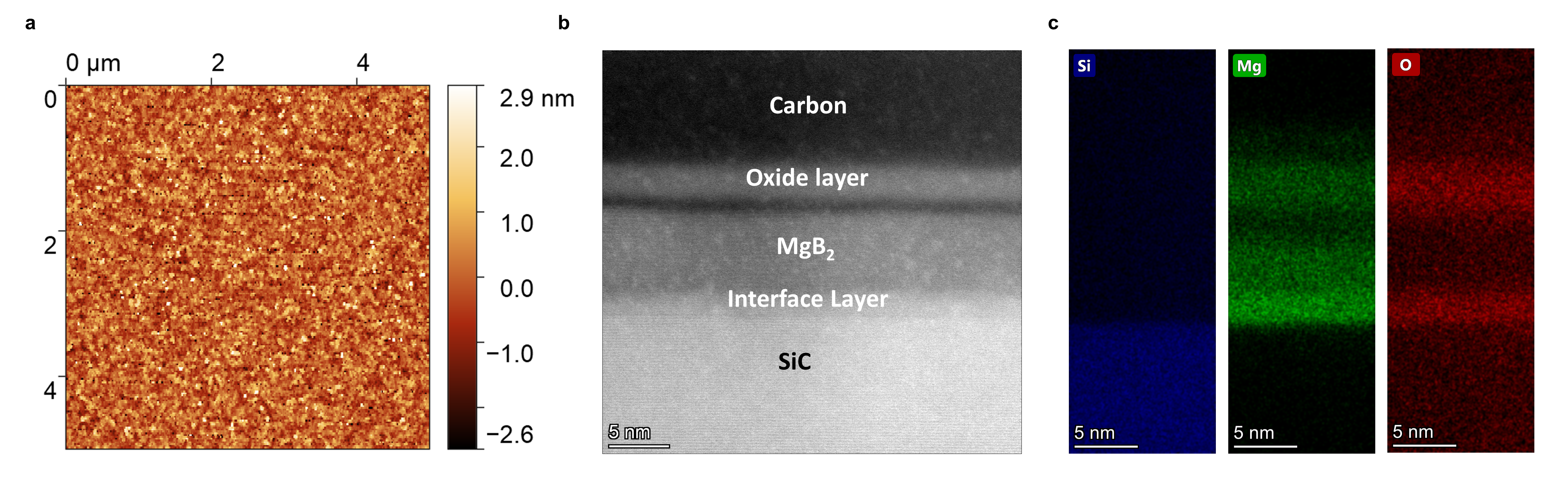}
	\caption{\textbf{Material analyses of superconducting MgB$_2$ films.} \textbf{a,} Atomic-force microscope (AFM) image of the surface topography of a MgB$_2$ film (sample F180, \SI{Supplementary Information, Fig 1}). The root-mean-square (RMS) surface roughness over several micrometers was less than 1\,nm. \textbf{b,} Cross-section of thin MgB$_2$ film taken via scanning transmission electron microscopy (STEM). Using focused-ion beam (FIB) preparation, a cross-section of the material was transferred onto a suitable TEM grid. The presence of carbon on top of the film is a result of the sample preparation. 
	\textbf{c,} Energy-Dispersive X-ray Spectroscopy (EDS) for elemental characterization of Si, Mg, and O. The EDS spectrum provides the compositional information for the deposited MgB$_2$ films. Notably, the superconducting MgB$_2$ core is sandwiched between two oxides. While the MgB$_2$ films readily pick up oxygen from the ambient atmosphere forming oxides on the top, the presence of O$_2$ at the interface is likely from residual SiO$_x$ on the SiC substrate surface.}
	\label{Fig2}
\end{figure*}

\textbf{Design and characterization of the MgB$_2$ detectors.}  To fabricate detectors out of MgB$_2$ films, we systematically optimized the Hybrid Physical-Chemical Vapor Deposition (HPCVD) process to produce relatively thin films with low surface roughness and high critical temperature. We concentrated on 5-10 nm thin films (Fig.~\ref{Fig2}b), where a high critical-current density can be achieved\cite{Cherednichenko_2021}. This thickness range also facilitates stronger absorption of incident light in the infrared range\cite{PhysRevB.80.054510}. Thicknesses were verified using x-ray reflectivity (XRR) (See \SI{Supplementary Information}) and scanning transmission electron microscopy (STEM). We also used cross-sectional energy dispersive x-ray spectroscopy (EDS) to identify and map elemental composition. Fig.~\ref{Fig2}c shows the results of qualitative analyses of a superconducting MgB$_2$ film deposited on the 6H-SiC substrate (See Methods and \SI{Supplementary Information} for details). While the total thickness of the studied sample was $\approx$12\,nm, the superconducting core of MgB$_2$ was measured to be only $\approx$7\,nm, and observed to be sandwiched between two oxide layers, one on the top of the film (2.5\,nm) and the other one at the substrate interface (1.5\,nm). A thin layer (assumed to be boron) was observed between superconducting MgB$_2$ and the top oxide layer. Boron is difficult to detect due to its light atomic mass but has been successfully identified in this region in similar samples (See Supplementary Information). These samples exhibited surprising resilience to degradation during standard lithography, etching, and contact deposition processes as well as storage in nitrogen over several months. Other films (unpatterned) lasted over three years in a vacuum desiccator. 
However, similar films have shown degradation under similar storage and processing conditions, indicating that the problem of the reliability and robustness of such films remains unresolved.

We analyzed the sample surface using Atomic Force Microscopy (AFM), observing a surface micro-roughness (RMS) of less than 1 nm over several square micrometers of film area (Fig.~\ref{Fig2}a). This low roughness suggests a minimal variation of the superconducting energy gap across the structure which is crucial for avoiding constrictions and thus maximizing the switching current\cite{cooper1961superconductivity}.

We used electron-beam lithography (EBL) to define a meander structure for SNSPDs with various dimensions, $1200-10200\mum$ in length and $1-5\mum$ in width, with a filling factor of 0.28 (Fig.~\ref{Fig1}a). For fabrication details and device geometries, see Methods and Supplementary Information. We used 12 magnesium diboride films with various thicknesses for our experiment. Out of these, six films underwent material analyses, while the remaining six were employed in the nanofabrication process to create detectors.

These devices, as initially fabricated, exhibited only dark counts. Motivated by our desire for single-photon detection we deliberately added defects by using 30-keV helium ion (He$^+$) irradiation \cite{PhysRevApplied.12.044040}. In contrast to patterning with heavier ions (e.g. gallium or xenon), the exposed regions were not observably etched by the He$^+$ beam irradiation. The detector area was entirely exposed with He$^+$ ions of $5\times10^{15}$ ions/cm$^2$ (See Supplementary Information). Such exposure had only a mild effect on switching current $I_\text{sw}$, $T_\text{c}$, and the normal state resistance $R_\text{s}$ (all changed by less than 5\%).

After fabrication, we characterized the transport properties of our high-$T_\text{C}$  superconducting MgB$_2$ microwires. Figure~\ref{Fig1}b shows the temperature dependence of the resistance, $R(T)$,  of a typical MgB$_2$ fabricated device, revealing the critical temperature of 37.6$\pm$0.3\,K as determined from the maximum of the $dR/dT(T)$. The obtained value is somewhat lower than those obtained for the parent MgB$_2$ film (38.7\,K). This decrease indicates a mild degradation of the material's superconducting properties during the fabrication and ion beam irradiation. Additional information on superconducting and electrical properties can be found in \SI{Supplementary Information}.

\begin{figure*}[ht!]
\centering\includegraphics[width=0.9\linewidth]{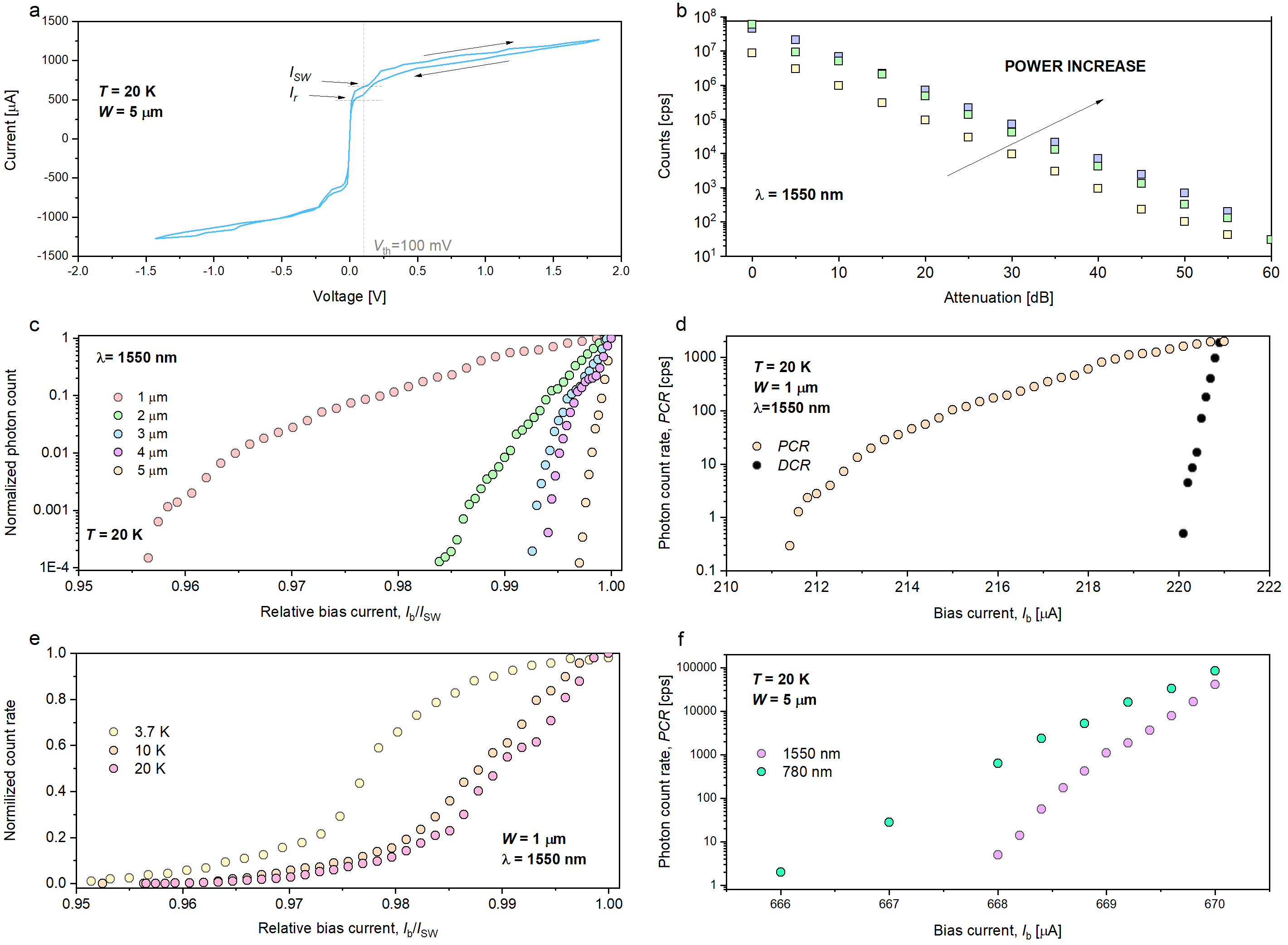}
	\caption{\textbf{Single-photon detection by superconducting MgB$_2$ microwires.} \textbf{a,} $I\mhyphen V$ curve for 5\,$\mu$m MgB$_2$ microwire detector measured at $T=20~$K in the two-terminal configuration.  \textbf{b,} Photon count rate vs attenuation factor, at different powers of incident lights for 1$\mu$m-wide MgB$_2$ device at given $I_\mathrm{b}$ $\approx$ 0.98$I_{sw}$ and $\lambda=1.5~\mu$m. Measurement shows linearity over five orders of magnitude in count rates.   \textbf{c,} Normalized \textit{PCR} vs relative bias current for different wire widths measured in the microscale MgB$_2$ devices at 20\,K.   \textbf{d,} Photon and dark-count rate ($DCR$) of 1-$\mu$m wide detector as a function of the absolute bias current ($I_b$) at 20\,K. \textbf{e,} The \textit{PCR}, normalized to its maximum value, as a function of the relative bias current, $I_\mathrm{b}/I_{sw}$, measured in 1-$\mu$m MgB$_2$ detector at given temperatures $T$ and 1550-nm wavelength ($\lambda$). Notably, the count rate at 3.7\,K is approaching a saturation plateau that suggests the internal detection efficiency of absorbed photons is approaching 100\%.  \textbf{f,} \textit{PCR} vs $I_\mathrm{b}$ for different $\lambda$ measured in 5-$\mu$m wide MgB$_2$ device at $T=20$~K.
 }
	\label{Fig3}
\end{figure*}

An important characteristic of single-photon detectors, enabling the generation of a voltage pulse upon single-photon absorption, is their metastable state that emerges under current biasing\cite{skocpol1974self}. This metastable state is essential for detector operation and appears due to competition between the current-induced Joule self-heating of the microwire in the resistive state and electron cooling processes. This state manifests as a hysteresis in the $I\mhyphen V$ characteristics. Figure~\ref{Fig3}a shows an example of an $I\mhyphen V$ curve measured in a 5 $\mu$m-wide MgB$_2$ microwire device at a temperature (20 \,K) well below $T_\text{c}$ when the microwire is current-biased. A slight hysteretic behavior, characterized by the switching current $I_{sw}=673~\mu$A and the retrapping current $I_r=490~\mu$A, is observed. 

We characterized the switching current of an array of devices by performing $I\mhyphen V$ sweeps for wire widths from $1\mum$ to $5\mum$ at different temperatures; the switching current results are collected in Supplementary Information. They do not follow linear fits to the wire width, as one would expect the approaching zero, likely due to exceeding the width of the wire in relation to the Pearl length\cite{pearl1964current} (this value is $1.55\mum$ for the penetration depth of 90 nm\cite{Cherednichenko_2021} ) and non-uniform current distribution in the superconducting portion\cite{pearl1964current}. Notably, the deviation from the linear trend becomes more pronounced as the temperature decreases, aligning with the temperature-dependent penetration depth.

\textbf{Photoresponse measurements.} To perform the photoresponse measurements, we mounted the chip with MgB$_2$ detectors in a variable-temperature cryostat equipped with RF coax cables and an optical fiber. The latter was held approximately 1\,cm away from the device so that a defocused continuous wave laser beam covered the whole chip area. The simplified circuit diagrams, used for the photoresponse measurements, can be found in [ see Figure 3b\cite{Charaev2023} ]. Detectors were measured in a conventional SNSPD configuration in which the device was biased through a DC input of the bias tee using an isolated voltage source with a bias resistor ranging from 1 to $10\,\text{k}\Omega$. The AC output was connected to the low-noise amplifier whose output was fed to an oscilloscope or a photon counter. To mitigate latching effects, we used a low-bandwidth reset loop formed by a shunt inductor, $L_{sh}$, and resistor, $R_{sh}$~\cite{RLshunt1} in parallel with the device.

Figure~\ref{Fig1}c shows an example of a generated photovoltage, measured across the current-biased MgB$_2$ microwires when the device was exposed to the laser beam radiation of wavelength $\lambda=1.5~\mu$m. The traces of these devices shared common features with the photoresponse of conventional NbN SNSPDs. After reaching the maximum value within $\sim1\,$ns, the voltage exhibited a slower decay with the characteristic time $\tau$, often referred to as dead or recovery time, which depends on the total kinetic inductance, $L_\mathrm{k}$, of the superconducting circuit and the load resistance\cite{kerman2009electrothermal}. The measured values (in particular, an example in Fig.~\ref{Fig1}c), $\tau\approx1.3~$ns (determined as the time when the signal dropped to 30$\%$ of its maximum value) are in agreement with previous measurements of the kinetic inductance in MgB$_2$ films for different detector lengths at various temperatures\cite{Cherednichenko_2021}. The voltage spikes in MgB$_2$ devices were observed below and above liquid-helium temperature and could be detected up to $T=20~$K. At higher temperatures, detectors did not exhibit a hysteresis, and thus no voltage pulses were observed upon illuminating the detectors with low-intensity laser light. 

\textbf{Single-photon sensitivity of MgB$_2$ detectors.} The MgB$_2$ devices exhibited single-photon sensitivity in the technologically-important 1550-nm telecommunications wavelength at the different powers of incident lights (Fig.~\ref{Fig3}b) and rates up to 100 Mcps. To obtain further insight into the performance of detectors, we recorded the photon count rate, \textit{PCR} (the number of pulses per unit time), as a function of the bias current, $I_\mathrm{b}$. In these measurements, the light underwent an additional 40 dB of attenuation. Figure~\ref{Fig3}c shows the \textit{PCR} normalized to its maximum value, measured in MgB$_2$ devices with widths of 1-$5\mum$ upon exposing it to the $\lambda=1.5~\mu$m laser light at 20\,K. Despite the biasing near the transition point at 99\% of the switching current, the operational bias range remains in the microampere range, making it similar to low-temperature detectors.
In the dark, spontaneous voltage pulses emerge when the device is biased close to the switching current (97-99\%) (Fig.~\ref{Fig3}d), similar to low-$T$ detectors. The absolute value of these dark counts did not exceed $2\times10^3$ s$^{-1}$, comparable to the values in conventional NbN SNSPDs. 
Upon illumination, the counts appeared at onset current $I_\mathrm{b}=0.96~I_\text{c}$ whereas the \textit{PCR} of the $1\mum$ wide device showed some tendency to saturation upon approaching $I_{sw}$ at 3.7\,K (Fig.~\ref{Fig3}e).
This saturation indicates the approaching a high internal detector efficiency \cite{marsili2013detecting}. With increasing $T$ to 20\,K, the onset current decreased together with $I_\mathrm{sw}$, as expected for superconducting devices(Fig.~\ref{Fig3}e).
Furthermore, we found that the \textit{PCR} for a given $I_\mathrm{b}$ differed for photon energies corresponding to photons with $\lambda=780~$nm and $\lambda=1.5~\mu$m (Fig.~\ref{Fig3}f). Single-photon operation was established through the observation of linear scaling of the photon count rate on the radiation power.  

We also found that the timing jitter of MgB$_2$ detectors is similar to that of conventional SNSPDs ($\sim$50 ps) (see Supplementary Information) with an active area at least one order of magnitude smaller than our devices. This experiment was performed using the apparatus as described in Supplementary Information.

\textbf{Discussion and outlook.} While the experimental results show single-photon detection in micro-scale wire MgB$_2$ detectors at elevated temperatures for the first time, a number of features of the data require discussion.
The key aspects to discuss are (1) the details of the $I\mhyphen V$ characteristics after He$^+$ exposure; (2) the mechanism of defect formation with He$^+$ ion exposure; (3) the impact of the two-band character of MgB$_2$; and (4) the mechanism of photodetection in microwires.

In Fig.~\ref{Fig3}, the IV-curve depicts multiple transitions occurring during the distinct jump from the superconducting to the resistive state. This observation on IV curves appeared after irradiation with He$^+$ ions and was unexpected. This data supports the hypothesis of non-uniform post-exposure with He$^+$ ions. The instability of the beam along with stitching error\cite{he2020helium} is expected to result in variation of the superconducting gap over the detector area. We observed, at various temperatures and low bias currents, a noticeable increase in photon count rate with bias current which supports this hypothesis, suggesting localized areas with a lower gap and thus relatively higher internal detection efficiency. We furthermore expect non-uniformity might have been introduced in the patterning and processing, therefore different regions of the wire could be participating in the detection process at different $T$ and $I_\mathrm{bias}$. With process development including He$^+$ irradiation dose distribution correction, improved results may thus be expected.

The role of He$^+$ irradiation at various doses on MgB$_2$ films remains not fully explored. Drawing from the data reported for low-$T_\text{C}$ SNSPDs composed of NbTiN\cite{He2023} and NbN\cite{PhysRevApplied.12.044040}, it's observed that even at He$^+$ dosages considerably lower than those necessary to induce noticeable effects on transport properties like critical current density or normal state conductivity, there is an enhancement in single-photon detection efficiency. In our study, we employed a dosage of $5\times10^{15}$ ions/cm$^2$ (equivalent to 50 ions/nm$^2$), a dose that aligns with the reported dosage required for 8-to-12\,nm-thick NbTiN films to begin exhibiting single-photon response. This dose resulted in only modest variations, up to 10\%, in parameters such as critical current density, critical temperature, and sheet resistance. These results closely approximate our findings. Nevertheless, comprehensive investigations in this area are still pending.

Apart from affecting MgB$_2$ film itself, the process of helium ion implantation causes defects and amorphization in the SiC substrate\cite{fan2023xenon}. As a result, the efficiency of energy transfer from the superconducting structure to the substrate is likely to be affected. This effect may prolong the lifespan of the normal domain, thereby increasing the probability of detection before the normal domain collapses. In devices subjected to a relatively low dose of irradiation ($5\times10^{15}$ ions/cm$^2$), we noticed a decrease in the retrapping current. While the switching current remained unchanged, the retrapping current reduced to 90\% of its nominal value compared to non-irradiated devices (Supplementary Information).

Although magnesium diboride has two bands with different gaps: $\pi$-band with a gap of 0.6$k_\text{B}$$T_\text{C}$ and $\sigma$-band with a gap of 2.2$k_\text{B}$$T_\text{C}$, the electron doping by atoms may initiate the interband scattering in MgB$_2$. The filling band effect increases the value of the small gap and, together with band filling, leads to the merging of the two gaps\cite{kortus2005band}. Our work points out that it is the uniformity of the superconducting gap, rather than the absolute gap, that is crucial.



It is also worth noting that, the width of wires surpassed the Pearl length in MgB$_2$. Wires wider than the Pearl length are expected to display non-uniform current distribution across the wire. The calculated Pearl length for MgB$_2$ films, approximately $1.3\mum$, is four times lower than the widest wire used in our experiment. As a result, the ratio of the experimentally observed switching current density to the current density for spontaneous vortex entry is expected to degrade, proportionally to the following: $\sim1/\sqrt{1+W/2\pi \Lambda_\text{P}}$\cite{PhysRevApplied.7.034014}, where $\Lambda_\text{P}$ is the Pearl Length. Surprisingly, contrary to theoretical predictions\cite{PhysRevApplied.7.034014}, we observed single-photon detection in devices with wire widths of up to $5\mum$. This observation challenges the previously known limit in wire width from a practical standpoint, even though the mechanism of detection under these conditions remains unclear.
\newpage
\indent 
\section*{Methods}
\subsection*{Sputtering and fab}
To fabricate detectors out of MgB$_2$ films we started with film deposition. Ultra-thin MgB$_2$ films were made using Hybrid Physical-Chemical Vapor Deposition (HPCVD) in a custom-built system\cite{zeng2003superconducting, novoselov2016study}. Magnesium melts at 650 degrees Celsius and interacts with boron. The latter is produced via the thermal decomposition of gaseous diborane (B$_2$H$_6$) supplied in a mixture with hydrogen. Films were made on (0001) SiC substrates which provide excellent lattice matching to the hexagonal lattice of c-oriented MgB$_2$ films.

Devices were fabricated out of MgB$_2$ films on the 6H-SiC substrate with 5 nm titanium/ 50 nm gold contact pads using standard e-beam lithography, followed by metal deposition and lift-off in $45^\circ$C N-Methylpyrrolidone (NMP). In order to produce the wires out of MgB$_2$ films, another round of e-beam lithography was performed, using 330-nm-thick high-resolution positive e-beam resist (ZEP520A). The patterns were then transferred onto the MgB$_2$ film by Ar$^+$ milling at a beam voltage of 300 V and Ar flow of 9 sccm for 12 minutes in a series of one-minute etches with three-minute-long intervals between each etch to avoid overheating of the sample. The residual resist was removed by immersing the substrate in the $45^\circ$C NMP.

After fabrication and initial testing, we irradiated our samples with He$^+$ using a Zeiss Orion Microscope equipped with a Raith pattern generator. The irradiation was realized by sweeping the beam across the detector area. The exposed area was a rectangle with dimensions of active detector areas (to cover the whole meander area). The beam-limiting aperture was set to the largest diameter to maximize the beam current.  The dose was varied in the range from 10$^{15}$ to 10$^{20}$ ions/cm$^2$ (\SI{Supplementary Information}). 

\subsection*{Scanning Transmission Electron Microscopy (STEM)}
STEM imaging was performed with a probe-corrected Thermo Fisher Scientific Themis Z G3 60–300\,K V S/TEM operated at 200\,K V with a beam current of 30-90 pA and 25 mrad convergence angle. Elemental analysis was accomplished by using EDS (Energy Dispersive X-ray Spectroscopy). For cross-sectional STEM characterization, samples were prepared using a  Raith VELION focused ion beam-scanning electron microscope (FIB-SEM) system with carbon and platinum protection layers on top of the sample.
\newline
\section*{Acknowledgments}

This work was carried out in part through the use of MIT.nano’s facilities. I. Charaev acknowledges support for this work from Brookhaven Science Associates, LLC award number 030814-00001. E. Batson acknowledges support from the National Science Foundation under Grant No. EEC-1941583 and from the National Science Foundation Graduate Research Fellowship under Grant No. 2141064. This work was performed in part on the Raith VELION FIB-SEM in the MIT.nano Characterization Facilities (Award: DMR2117609). K.R. acknowledges funding and support from a MIT MathWorks Engineering Fellowship and ExxonMobil Research and Engineering Company through the MIT Energy Initiative. J.D.T. acknowledges support from the Independent Research Fund Denmark through Grant 9035-00006B. We also thank Boris Korzh (JPL) and Frances M. Ross for the helpful discussions. The work at Chalmers University of Technology was carried out with a support from Swedish Research Coun cil (2019-04345) and Swedish National Space Agency (198/16).


\section*{Author contributions}
I.C, S.C, and K.K.B. conceived and designed the project. I.C., E.B., S.L.A. and M.C. performed transport and photoresponse measurements. I.C. and E.B. fabricated devices. F. I. performed helium ion irradiation of some devices. I.C., S.C., E.B., and K.K.B. analyzed the experimental data. K.R. and Y.Y. performed STEM measurements. J.T. performed AFM measurements. S.C. deposited MgB$_2$ films.  V.D. fabricated nanowires for transport measurements. I.C. wrote the manuscript with input from all co-authors. K.K.B. and S.C. supervised the project. All authors contributed to discussions.

\section*{Competing interests}
The authors declare no competing interests.
\newline

*Correspondence to: ilya.charaev@physik.uzh.ch, berggren@mit.edu, serguei@chalmers.se

\bibliographystyle{naturemag}
\bibliography{Bibliography.bib}

\end{document}


\title{Supplementary materials for \\ Single-photon detection using large-scale high-temperature MgB$_2$ sensors at 20 K}

\author{I. Charaev$^{*}$}
\affiliation{Massachusetts Institute of Technology, Cambridge, MA 02139, USA}
\affiliation{University of Zurich, Zurich 8057, Switzerland}
\author{E. K. Batson}
\affiliation{Massachusetts Institute of Technology, Cambridge, MA 02139, USA}
\author{S. Cherednichenko$^{*}$}
\affiliation{Department of Microtechnology and Nanoscience, Chalmers University of Technology, Göteborg SE-41296, Sweden}
\author{K. Reidy}
\affiliation{Massachusetts Institute of Technology, Cambridge, MA 02139, USA}
\author{V. Drakinskiy}
\affiliation{Department of Microtechnology and Nanoscience, Chalmers University of Technology, Göteborg SE-41296, Sweden}
\author{Y. Yu}
\affiliation{Raith America, Inc., 300 Jordan Road, Troy, NY 12180}
\author{S. Lara-Avila}
\affiliation{Department of Microtechnology and Nanoscience, Chalmers University of Technology, Göteborg SE-41296, Sweden}
\author{J. D. Thomsen}
\affiliation{Massachusetts Institute of Technology, Cambridge, MA 02139, USA}
\author{M. Colangelo}
\affiliation{Massachusetts Institute of Technology, Cambridge, MA 02139, USA}
\author{F. Incalza}
\affiliation{Massachusetts Institute of Technology, Cambridge, MA 02139, USA}
\author{K. Ilin}
\affiliation{Institute of Micro- and Nanoelectronic Systems, Karlsruhe Institute of Technology (KIT), 76187 Karlsruhe, Germany}
\author{A. Schilling}
\affiliation{University of Zurich, Zurich 8057, Switzerland}
\author{K. K. Berggren*}
\affiliation{Massachusetts Institute of Technology, Cambridge, MA 02139, USA}

\maketitle

\setcounter{figure}{0}
\setcounter{section}{0}
\setcounter{equation}{0}
\renewcommand{\thesection}{}
\renewcommand{\thesubsection}{\arabic{subsection}}
\renewcommand{\thesubsubsection}{\arabic{subsection}.\arabic{subsubsection}}
\renewcommand{\theequation} {S\arabic{equation}}
\renewcommand{\thefigure} {S\arabic{figure}}
\renewcommand{\thetable} {S\arabic{table}}

\subsection{\textbf{
Superconducting MgB$_2$ film characterization}}
\begin{figure*}[ht!]
	\centering\includegraphics[width=0.7\linewidth]{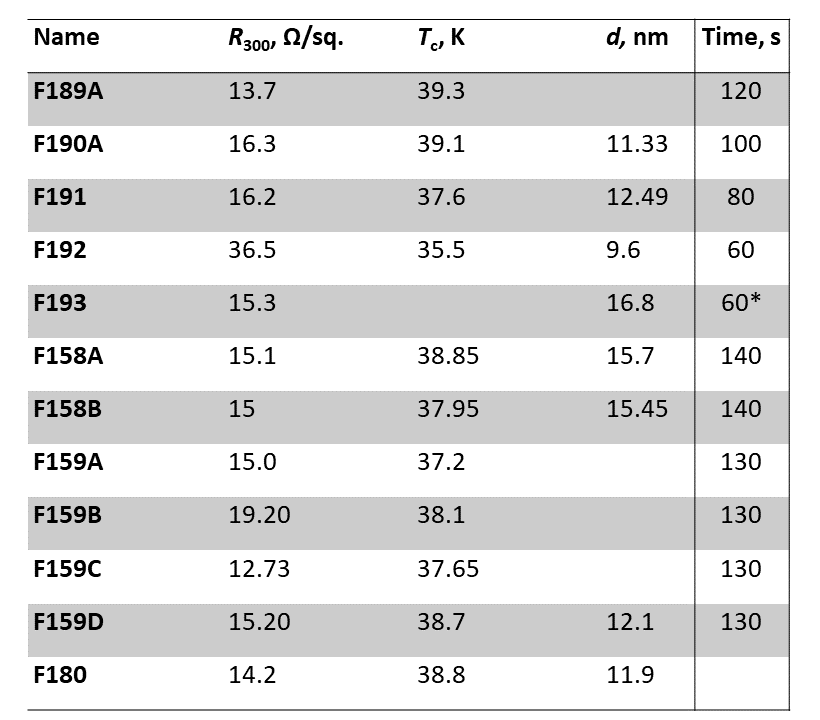}
	\caption{\textbf{Superconducting and transport properties of magnesium diboride films} where $R_\mathrm{300}$ is a sheet resistance at 300 K; $T_c$ - the critical temperature; $d$ - the thickness; Time - the deposition time.}
	\label{Films}
\end{figure*}

In total, we have deposited more than a dozen films in order to obtain MgB$_2$ SNSPD devices with the highest critical temperature. To reach this goal, we completed a variety of tasks such as (1) optimization of deposition temperature, (2) gaseous diborane B$_2$H$_6$ flow variation, (3) sweeping of process power, and (4) reduction of the deposition rate. In Fig.~\ref{Films} we summarize the superconducting and transport properties of deposited films. The sheet resistance was measured using a 4-probe technique. The films were characterized in terms of their RT-dependencies in closed-cycle cryostat from 300 down to 3.7 K. The critical temperature for this dataset was recorded at the temperature where $R_\mathrm{50K}$/2. The thickness of deposited films was controlled by XRR reflectivity (details below).

\subsection{\textbf{
Second critical magnetic field}}
\begin{figure*}[ht!]
	\centering\includegraphics[width=0.7\linewidth]{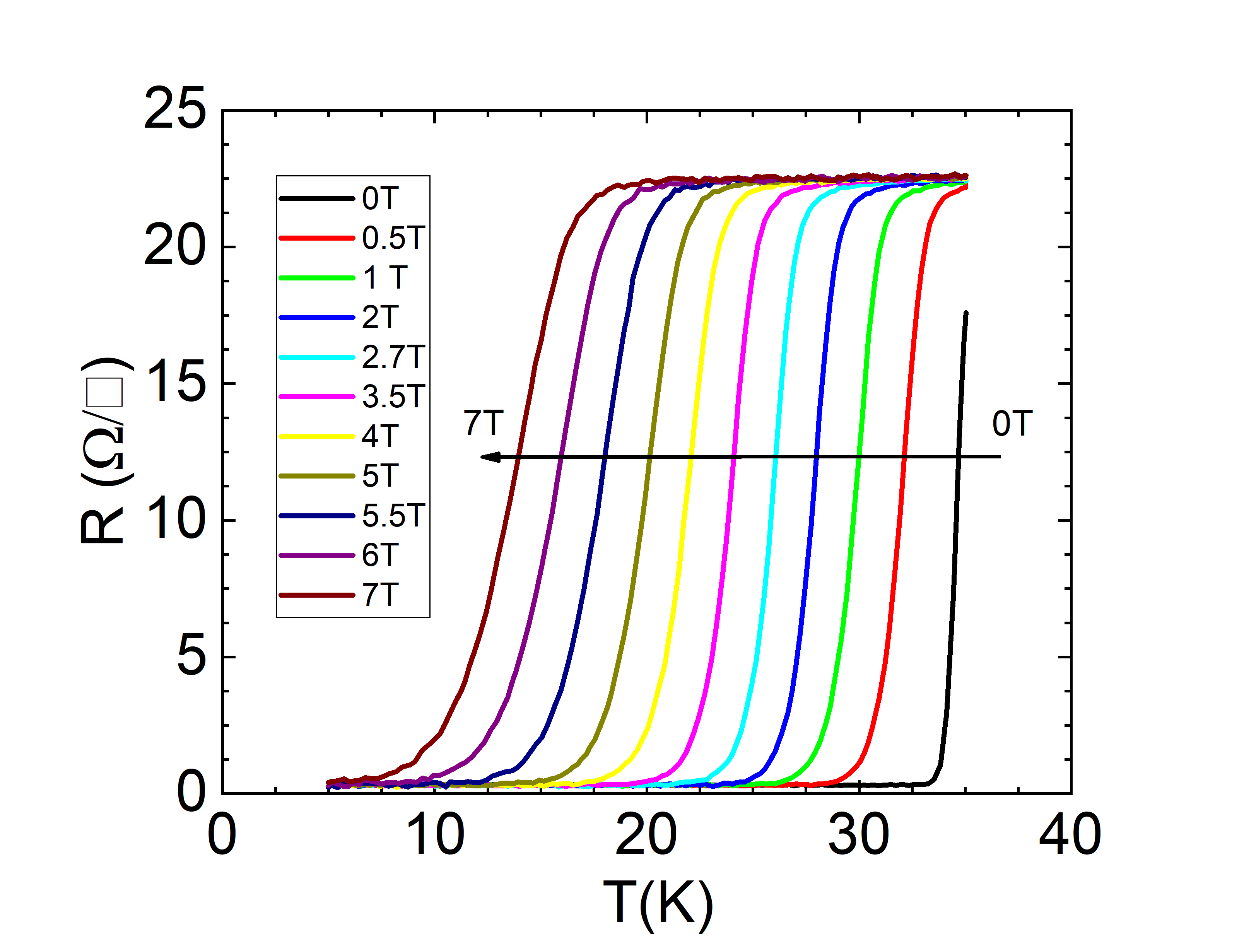}
	\caption{\textbf{Second critical magnetic field in magnesium diboride films}}
	\label{Bc2}
\end{figure*}

The temperature dependence of the second critical magnetic field $B_\mathrm{c2}$ was measured by applying an external magnetic field perpendicularly to the film surface. 
Calculation of the electron diffusion coefficient was accomplished by taking the linear part of the temperature dependence (Fig.~\ref{Bc2}) of $B_\mathrm{c2}$. This resulted in the value of the electron diffusion coefficient $D$ = 0.98 cm$^2$/sec.

\begin{figure*}[ht!]
	\centering\includegraphics[width=0.55\linewidth]{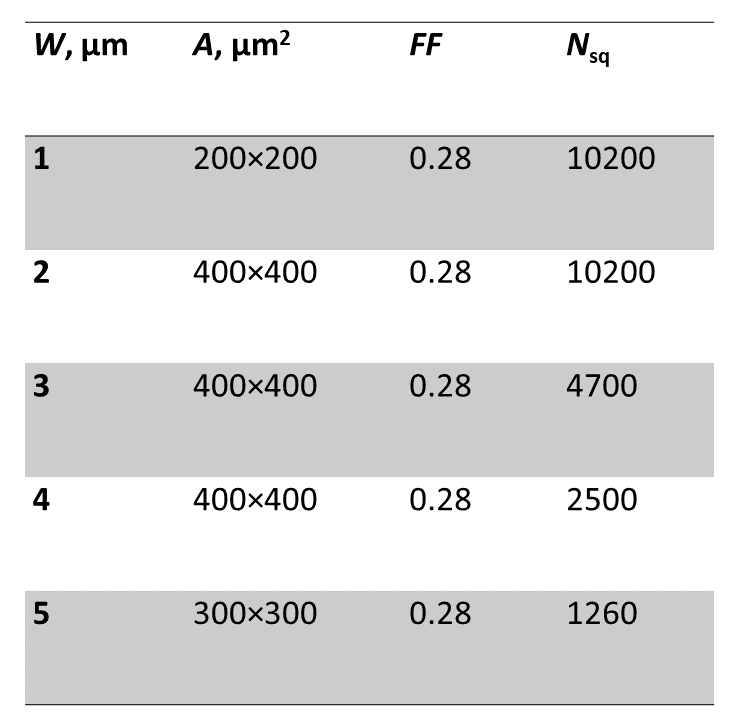}
	\caption{\textbf{Summary of microscale wide detector geometry.} $W$ is a width of wires; $A$ - the active area of detector; $FF$ - filling factor; $N_\mathrm{sq}$ - the number of squares.}
	\label{Geometry}
\end{figure*}

\subsection{\textbf{
Detector geometry}}
We designed the detectors with various widths and active areas. Fig.~\ref{Geometry} shows the summary of detector geometries. While the filling factor was kept the same to prevent current crowding in turns of meander, the active area and the number of squares varied in the range up to 400$\times$400~$\mu$m$^2$ and up to 10200 squares respectively.

\subsection{\textbf{
Switching current}}
\begin{figure*}[ht!]
	\centering\includegraphics[width=0.7\linewidth]{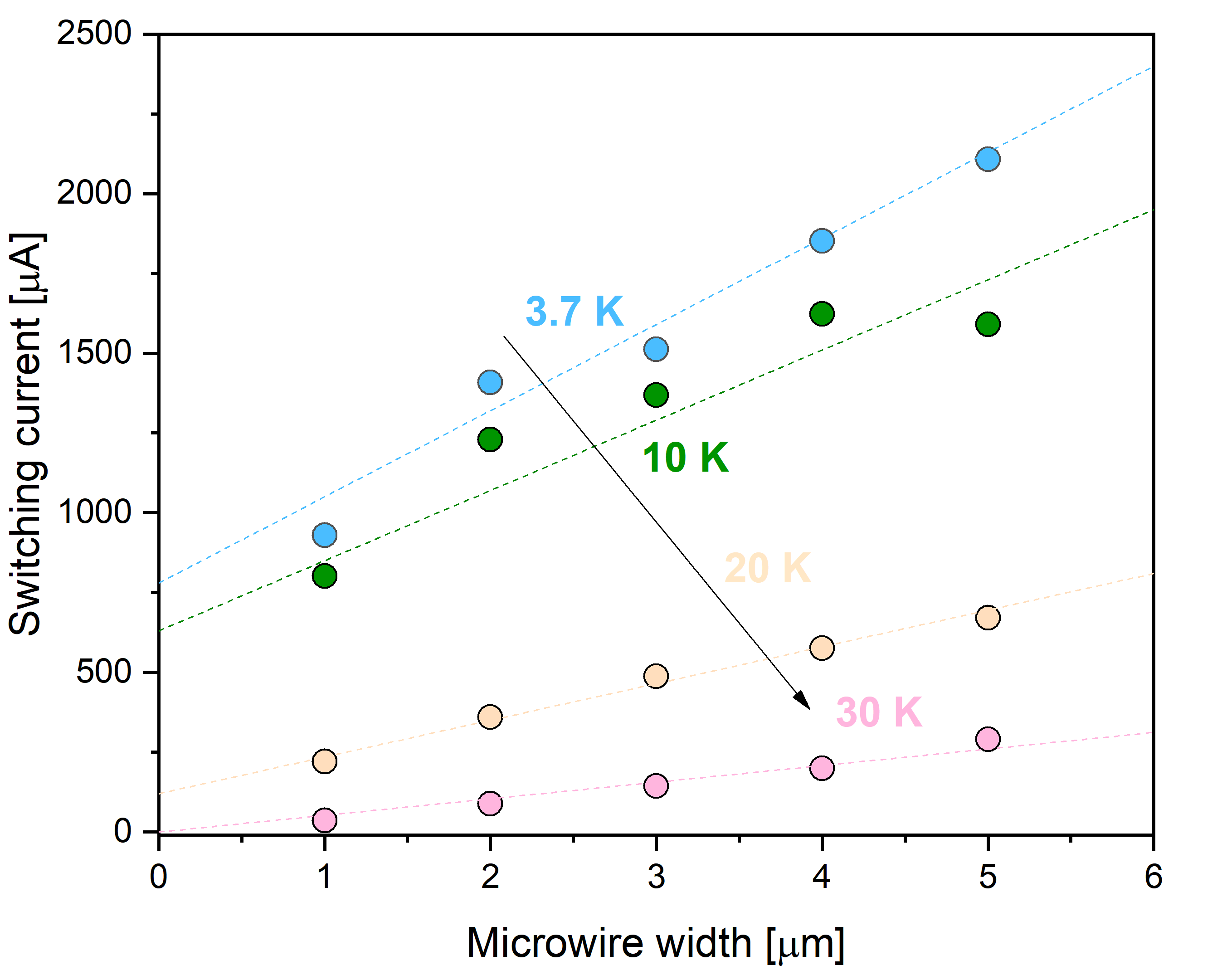}
	\caption{\textbf{Switching current} The dependence of the switching current on the device width and a bath temperature.}
	\label{Current}
\end{figure*}

We conducted an analysis of the switching current in devices, which involved performing I-V sweeps for wire widths ranging from 1 to 5 $~\mu$m at varying temperatures. The results of the switching current were plotted in Fig.~\ref{Current}. In general, the switching current does not exhibit a linear relationship with wire width. 

\subsection{\textbf{
He$^+$ dose tests for MgB$_2$ detectors}}

\begin{figure*}[ht!]
	\centering\includegraphics[width=0.6\linewidth]{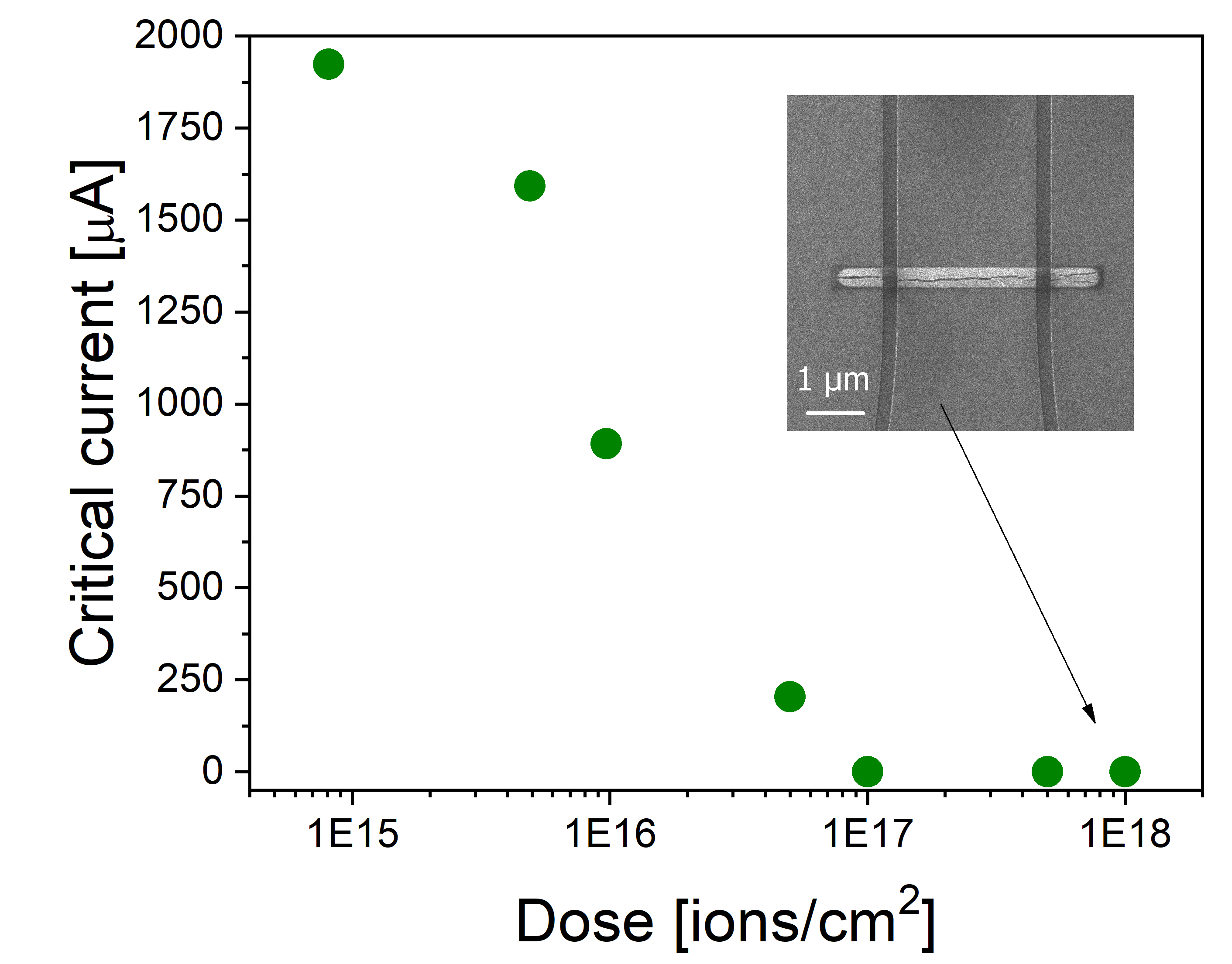}
	\caption{\textbf{He$^+$ dose test for MgB$_2$ microbridges.} Inset shows the SEM image of the exposed microbridge at the dose of $10^{18}$ ions/cm$^{-2}$.} 
	\label{Dose_He}
\end{figure*}

\begin{figure*}[ht!]
	\centering\includegraphics[width=0.9\linewidth]{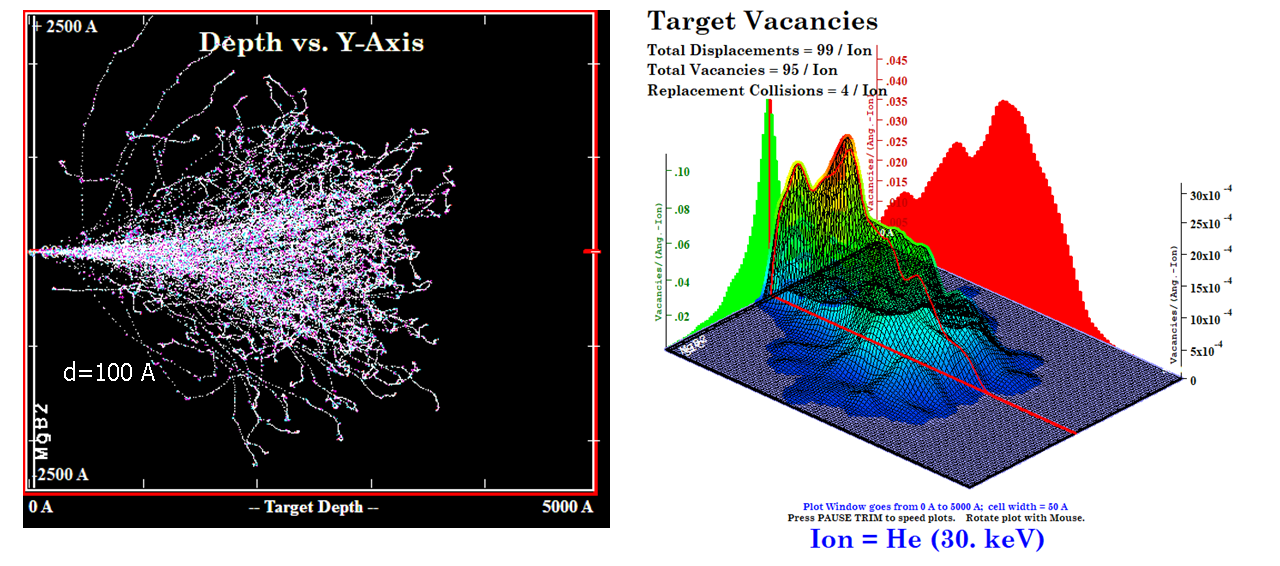}
	\caption{\textbf{Simulation of the interaction of 30\,keV helium ions with the MgB$_2$ device.} The results of the Monte-Carlo simulation performed using Stopping and range of ions in matter package (SRIM).}
	\label{SRIM.png}
\end{figure*}

\begin{figure*}[ht!]
	\centering\includegraphics[width=0.7\linewidth]{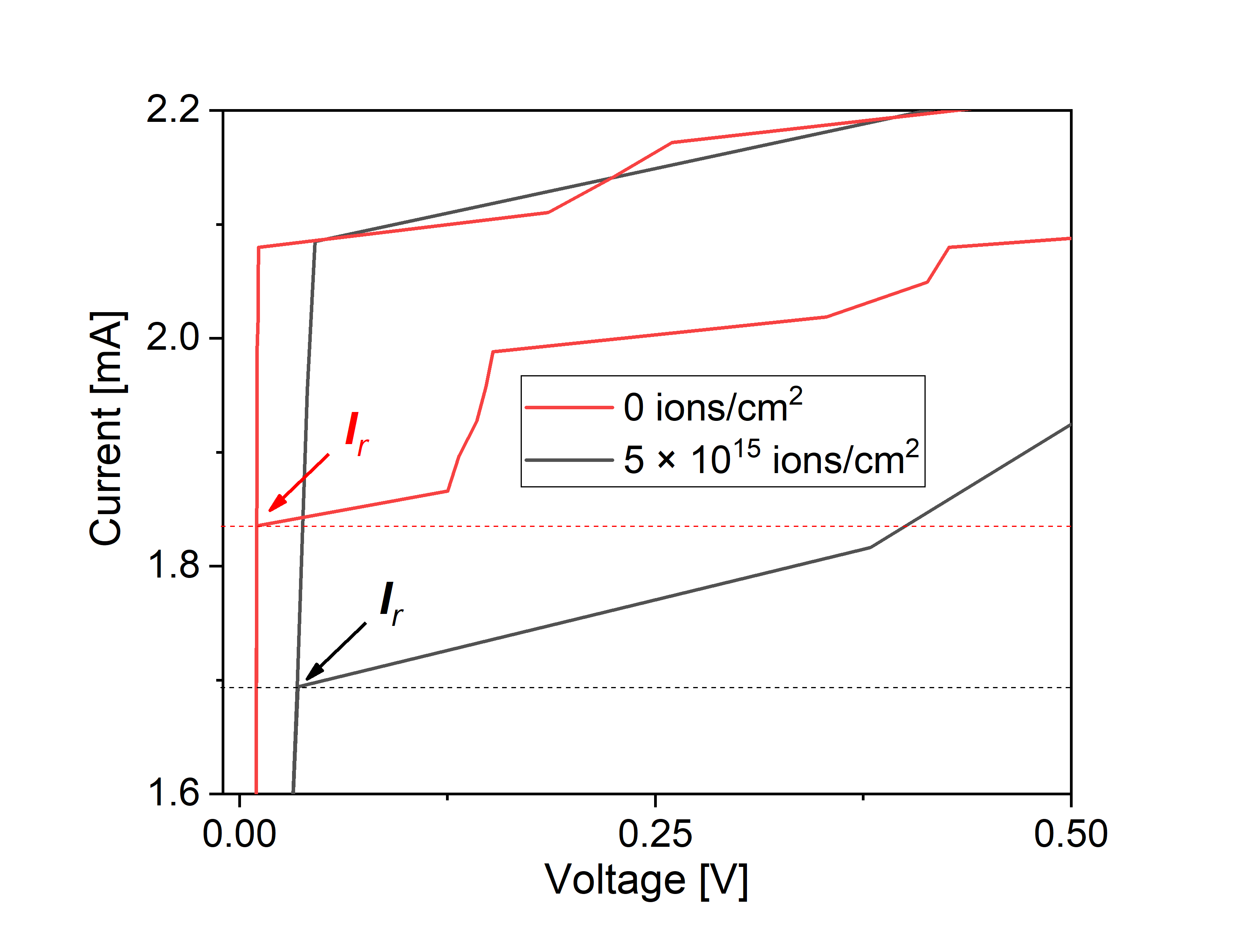}
	\caption{\textbf{ Effect He$^+$ irradiation on switiching and retrapping current.} Current-voltage characteristics taken from 5-$\mu$m wide MgB$2$ device before and after He$^+$ irradiation at 3.7\,K.} 
	\label{IV_RETR}
\end{figure*}

Before the fabrication of SNSPD out of MgB$_2$, we analyzed the He$^+$ beam exposure effects on the superconducting properties of the MgB$_2$-based films. To this end, we measured the critical current, $I_c$, of MgB$_2$ bridges as a function of the He$^+$ dose (Fig.~\ref{Dose_He}). We found that while $10^{15}$ ions/cm$^{2}$ had practically no effect on $I_c$, $10^{17}$ ions/cm$^{2}$ resulted in the full suppression of the superconductivity in this material. The dose $5\times10^{15}$ ions/cm$^{2}$ caused only mild suppression of $I_c$ but decreases of $I_r$ (Fig.~\ref{IV_RETR}) and thus to the enhancement of single-photon sensitivity of MgB$_2$ detectors. Additionally, we implemented a Monte-Carlo simulation of the interaction of helium ions with MgB$_2$ using SRIM (Fig.~\ref{SRIM.png}).


\subsection{\textbf{
X-ray reflectivity measurements}}

\begin{figure*}[ht!]
	\centering\includegraphics[width=0.6\linewidth]{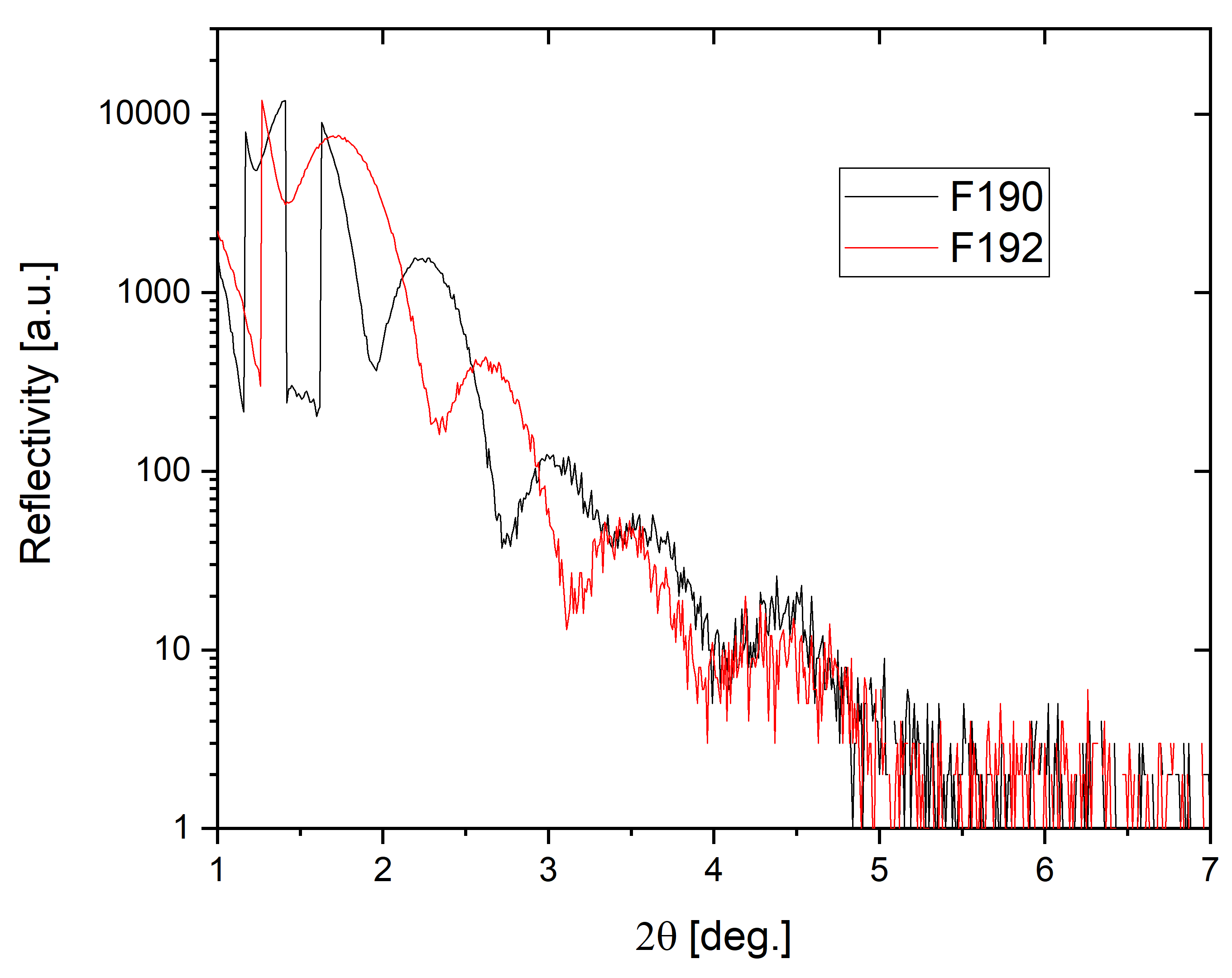}
	\caption{\textbf{X-ray reflectivity measurements.} Experimental X-ray reflectivity (XRR) as a function of twice the incident angle for films shown in the legend. }
	\label{XRR}
\end{figure*}

To verify the thickness in our MgB$_2$ films, we performed the XRR measurements using Rigaku Smartlab with an incident-beam Ge(022) monochromator, which is used for high-resolution XRR on relatively thick films up to ~300\,nm. The sample was precisely aligned and the measurement conditions were optimized automatically based on sample information. Fig.~\ref{XRR} shows a few examples of taken data for MgB$_2$ films.

\subsection{\textbf{
Energy Dispersive X-Ray Spectroscopy (EDS)}}


The material analysis was performed on a film prepared by the same growth process described in Methods.  The film showed a magnesium oxide on either side of the MgB$_2$ film (which is 2-6\,nm thick for each layer).

\begin{figure*}[ht!]
	\centering\includegraphics[width=0.8\linewidth]{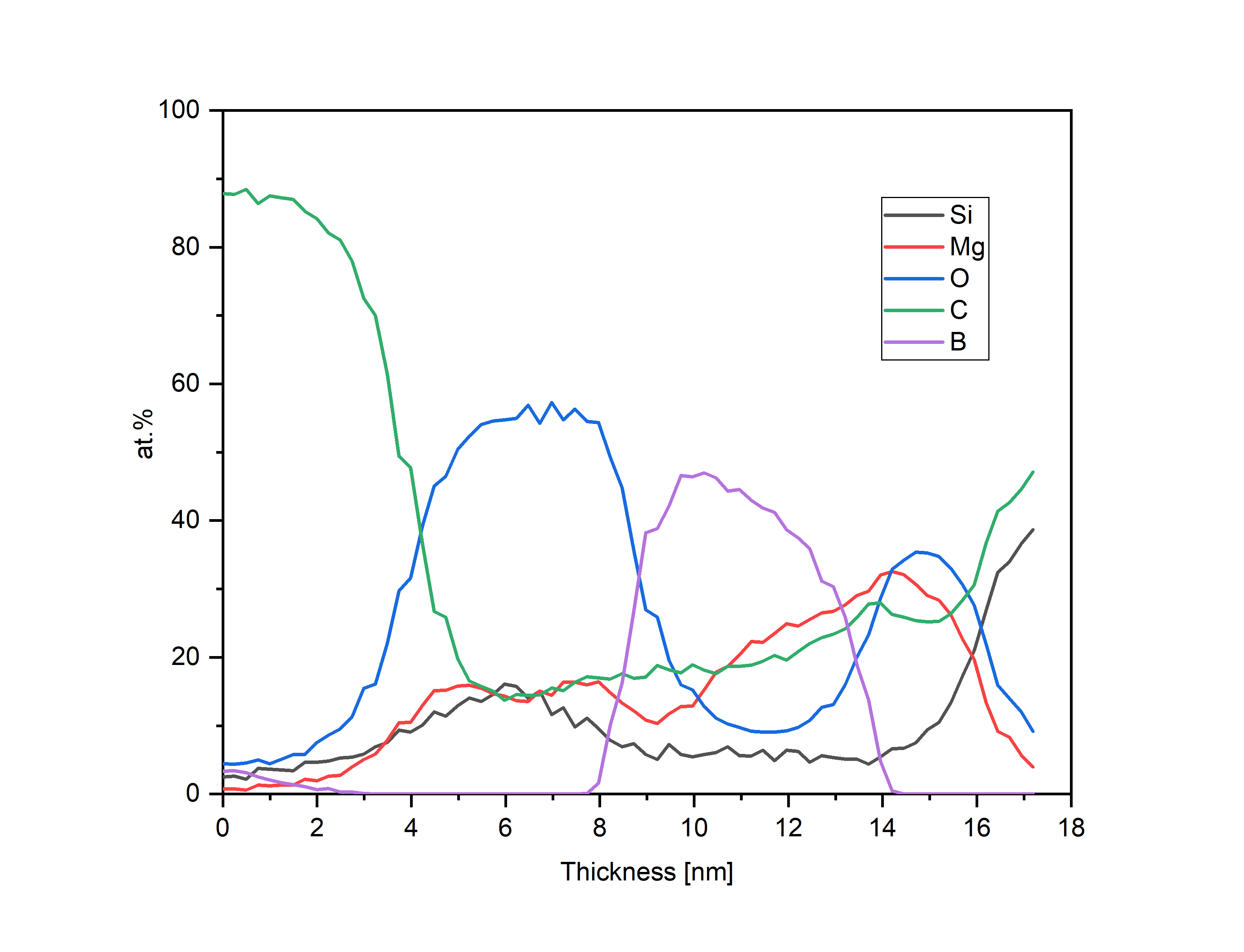}
	\caption{\textbf{EELS measurements performed on MgB$_2$ film prepared by the same growth process as used for the patterned films.} The left-hand side of the graph ($0-2$\,nm) corresponds to the top of the film with a carbon protection layer from FIB processing, while the right-hand side ($16-18$\,nm) shows the SiC substrate. The EDS atomic \% obtained are not quantitatively accurate for light elements such as B and O, however, are intended to give a qualitative overview of the elemental distribution across the sample thickness.}  
	\label{EDS}
\end{figure*}

\subsection{\textbf{
Timing jitter}}

\begin{figure*}[ht!]
	\centering\includegraphics[width=0.8\linewidth]{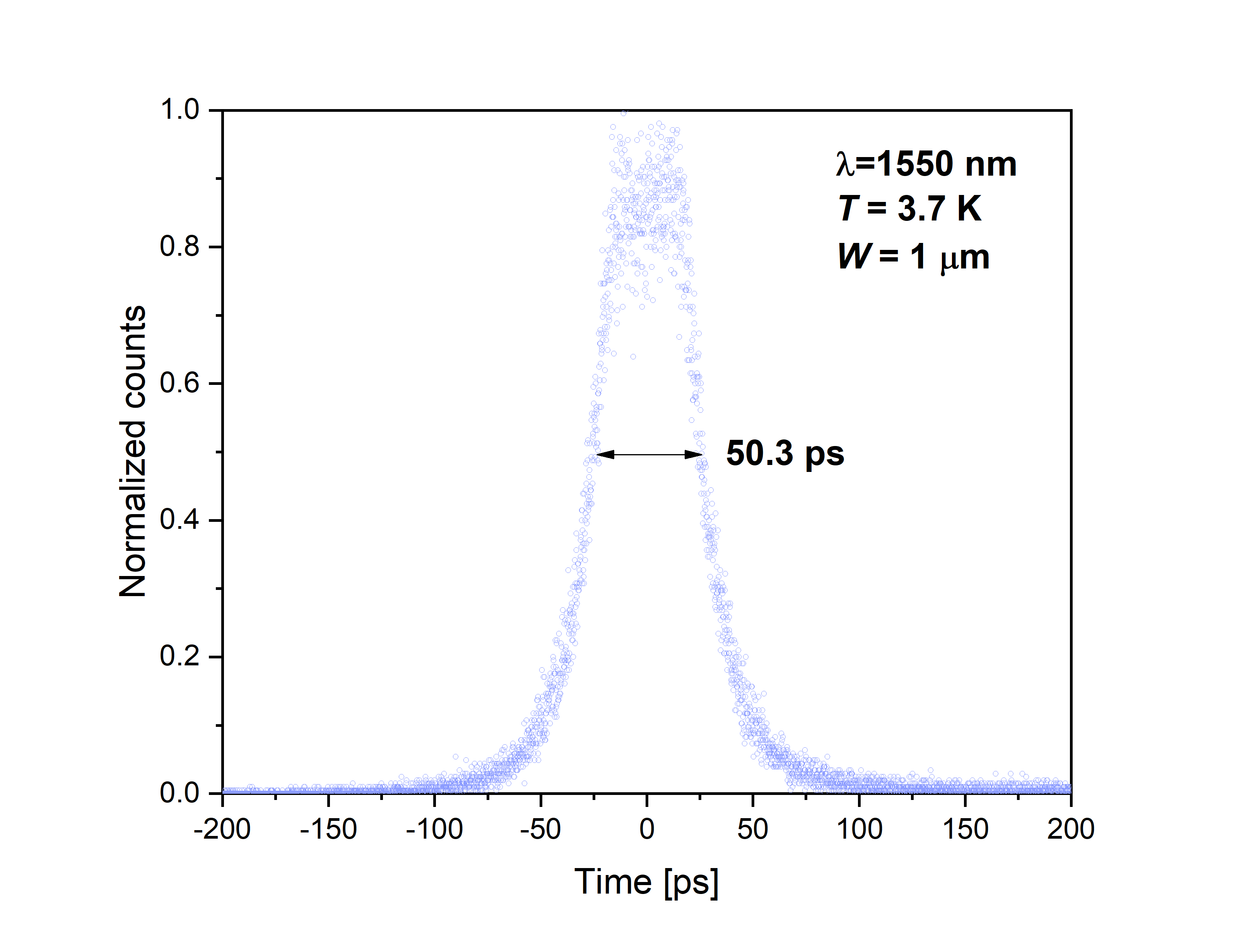}
	\caption{\textbf{Timing jitter measurements.} The combination of the detector's normalized instrument response function and the femtosecond laser synchronization signal is represented in a histogram. The jitter of 50.3 ps is indicated by arrows with a sign. }
	\label{Jitter}
\end{figure*}

To measure timing jitter, we used an experimental setup with  a 1550 nm femtosecond-pulsed laser that was adopted as the single-photon source. The real-time oscilloscope builds a statistical distribution of the arrival times of photon counting pulses. The distribution typically has an almost Gaussian profile
(Fig.~\ref{Jitter}). We defined the timing jitter as the full width at half maximum (FWHM) of the distribution. We found that the jitter of micro-scale MgB$_2$ detectors is similar to conventional SNSPD detectors to be 50.3 ps.

\bibliography{Bibliography.bib}

\clearpage
\newpage